\documentclass[12pt,twoside,english]{article}
\usepackage[T1]{fontenc}
\usepackage[latin9]{inputenc}
\usepackage[a4paper]{geometry}
\usepackage{graphicx}
\usepackage{esint}

\providecommand{\tabularnewline}{\\}

\newcommand{\lyxaddress}[1]{
\par {\raggedright #1
\vspace{1.4em}
\noindent\par}
}

\usepackage{babel}

\begin{document}

\title{Form factors in the presence of integrable defects}

\author{Zoltán Bajnok$^{a}$ and Omar el Deeb$^{b}$}

\maketitle

\lyxaddress{\begin{center}
\emph{$^{a}$Theoretical Physics Research Group of the Hungarian
Academy of Sciences,}\\
\emph{$^{b}$Institute for Theoretical Physics, Roland Eötvös University,
}\\
\emph{H-1117 P\'azm\'any s. 1/A, Budapest, Hungary}
\par\end{center}}

\vspace{-8cm}

\hspace{11cm}ITP-Budapest Report 645

\vspace{8cm}
\begin{abstract}
Form factor axioms are derived in two dimensional integrable defect
theories for matrix elements of operators localized both in the bulk
and on the defect. The form factors of bulk operators are expressed
in terms of the bulk form factors and the transmission factor. The
structure of the form factors of defect operators is established in
general, and explicitly calculated in particular, for the free boson
and for some operator of the Lee-Yang model. Fusion method is also
presented to generate boundary form factor solutions for a fused boundary
from the known unfused ones. 
\end{abstract}

\section{Introduction}

The bootstrap program aims to classify and explictly solve 1+1 dimensional
integrable quantum field theories by constructing all of their Wightman
functions (see \cite{FFBootstrap} for a recent review and references
therein). In the first step, called the S-matrix bootstrap, the scattering
matrix, connecting asymptotic in and out states, is determined from
its properties such as factorizability, unitarity, crossing symmetry
and Yang-Baxter equation (YBE) supplemented by the maximal analyticity
assumption \cite{exactS}. In the second step, called the form factor
bootstrap, matrix elements of local operators between asymptotic states
are computed using their analyticity properties originating from the
already computed S-matrix. Supposing maximal analyticity leads to
a set of solutions each of which corresponds to a local operator of
the theory \cite{FFbook}. In the third step these \emph{bulk form
factors} are used to build up the correlation (Wightman) functions
via their spectral representations and describe the theory completely
off mass shell. This program has been implemented for a wide range
of theories (see \cite{FFBootstrap}). . 

The analogous bootstrap program for 1+1 dimensional integrable \emph{boundary}
quantum field theories has been already developed. The first step
is called the R-matrix bootstrap \cite{GZ}: In boundary theories
the asymptotic states are connected by the reflection (R)-matrix,
which, as a consequence of integrability, factorizes and satisfies
the unitarity, boundary crossing unitarity and the boundary YBE (BYBE)
requirements. These equations supplemented by the maximal analyticity
assumptions make possible to determine the reflection matrices and
provide the complete information about the theory on mass shell. In
the second step we are interested in the matrix elements of local
operators localized both in the bulk and also at the boundary. Due
to the absence of translational invariance the \emph{bulk operators}'
one point functions acquire nontrivial space dependence which can
be calculated in the crossed channel using the knowledge of the boundary
state together with the bulk form factors \cite{1ptf}. For the matrix
elements of local \emph{boundary operators} axioms can be derived
from their analytical properties originating from the already computed
R-matrix \cite{BFF}. Supposing maximal analyticity leads to a set
of solutions each of which corresponds to a local boundary operator
of the theory and is uniquely related to a vector in the ultraviolet
Hilbert space. The explicit form of the boundary form factors determine
the boundary correlation functions via their spectral representation.
This provides a partial description of the theory off the mass shell
as a full description would include correlation functions of operators
localized in the bulk as well, but this complicated problem has not
been addressed yet.

Since any two dimensional \emph{defect} theory can be mapped to a
boundary theory \cite{DefBound} the development of a separate bootstrap
program for their solution seems to be redundant. However, integrable
defect theories are severely restricted and one can go much beyond
the boundary bootstrap program explained above: We can determine the
form factors of both types of operators, those localized in the bulk
and also the ones localized on the defect. With the help of these
form factors we are able to derive spectral representation for any
correlation function and in principle fully solve the theory off the
mass shell as we will show in the present paper. 

In developing a defect form factor progam the first step is the T-matrix
bootstrap. Interacting integrable defect theories are purely transmitting
\cite{DMS1,DMS2,Defboot} and topological. As a consequence a momentum
like quantity is conserved \cite{ClDef,DAT} and the location of the
defect can be changed without affecting the spectrum of the theory
\cite{BS,XXZ}. This fact, together with integrability leads to the
factorization of scattering amplitudes into the product of pairwise
scatterings and individual transmissions and enables one to determine
the transmission factors from defect YBE (DYBE), unitarity and defect
crossing unitarity \cite{KL,QAT,sgdef}. The second step is the defect
form factor bootstrap: Once the transmission factors are known we
can formulate the axioms that have to be satisfied by the matrix elements
of local defect operators. We will analyze operators localized both
in the bulk and also on the defect. By finding their solutions the
spectral representation of any correlator can be determined and the
theory can be solved completely. For simplicity we restricted our
interest in the paper for theories with a single particle type. 

The paper is organized as follows: In section 2 we introduce asymptotic
states in defect theories and the notion of the transmission matrix.
Then the coordinate dependence of defect form factors is determined.
By specifying the boundary form factor axioms we postulate the axioms
for diagonal defect theories. In section 3 we determine the form factors
of any operator localized in the bulk in terms of the transmission
factor and the bulk form factors. For operators localized on the defect
a procedure to calculate the general form factor solution is outlined.
In section 4 we apply this technology to determine the defect form
factors of the free boson and the Lee-Yang model. Finally we present
how the fusion method can be adapted to generate boundary form factor
solutions for a fused boundary from the known unfused ones.

\section{Defect form factors }

In this section we present the axioms for the matrix elements of local
operators between asymptotic states. To shorthen the discussion we
introduce Zamolodchikov-Faddeev (ZF) operators in order to describe
both the multiparticle transmission process as well as the properties
of the form factors.

\subsection{Asymptotic states and transmission matrix}

Multi-particle asymptotic states in integrable \emph{bulk} theories
can be formulated in terms of the ZF operators as \[
\vert\theta_{1},\dots,\theta_{n}\rangle=A^{+}(\theta_{1})\dots A^{+}(\theta_{n})\vert0\rangle\]
All particles have different momenta $p_{i}=m\sinh\theta_{i}$, thus
in the remote past they are not interacting and form an initial state
$\theta_{1}>\dots>\theta_{n}$. When time evolves they approach each
other and after the consequtive scatterings they rearrange themselves
into the opposite order:\[
\vert\theta_{1},\dots,\theta_{n}\rangle=\prod_{i<j}S(\theta_{i}-\theta_{j})\vert\theta_{n},\dots,\theta_{1}\rangle\]
Here $S(\theta)$ is the two particle scattering matrix which satisfies
unitarity and crossing symmetry \[
S(-\theta)=S^{-1}(\theta)\quad;\qquad S(i\pi-\theta)=S(\theta)\]
This multi-particle scattering process easily formulated with the
ZF algebra: 

\begin{equation}
A^{+}(\theta_{1})A^{+}(\theta_{2})=S(\theta_{1}-\theta_{2})A^{+}(\theta_{2})A^{+}(\theta_{1})+2\pi\delta(\theta_{1}-\theta_{2}-i\pi)\label{eq:ZFalgebra}\end{equation}
where we extended their definition for imaginary $\theta$ by postulating
the crossing property \begin{equation}
A(\theta)=A^{+}(\theta+i\pi)\label{eq:crossing}\end{equation}
see \cite{BFF} for the details. 

Once defects are introduced we have to make a distinction whether
the particle arrives from the left ($A$) or from the right ($B$)
to the defect. These particles can be even different from each other
as they live in different subsystems. A multiparticle state is then
described by \[
\vert\theta_{1},\dots,\theta_{n};\theta_{n+1},\dots,\theta_{m}\rangle=A^{+}(\theta_{1})\dots A^{+}(\theta_{n})D^{+}B^{+}(\theta_{n+1})\dots B^{+}(\theta_{m})\vert0\rangle\]
where the ZF operators $B^{+}$ create particles on the right of the
defect and satisfy similar defining relations to (\ref{eq:ZFalgebra})
with a possibly different scattering matrix. Yet, for simplicity,
we restrict our discussion to the case when the two subsystems are
identical with the same scattering matrix. Observe however, that this
does not imply space parity invariance, since the defect may break
it. In the initial state rapidities are ordered as $\theta_{1}>\dots>\theta_{n}>0>\theta_{n+1}>\dots>\theta_{m}$.
The final state, in which all scatterings and transmissions are already
terminated, can be expressed in terms of the initial state via the
multiparticle transmission matrix. \[
\vert\theta_{1},\dots,\theta_{n};\theta_{n+1},\dots,\theta_{m}\rangle=\prod_{i<j}S(\theta_{i}-\theta_{j})\prod_{i=1}^{n}T_{-}(\theta_{i})\prod_{i=n+1}^{m}T_{+}(-\theta_{i})\vert\theta_{m},\dots,\theta_{n+1};\theta_{n},\dots,\theta_{1}\rangle\]
Due to integrability it factorizes into pairwise scatterings and individual
transmissions: $T_{-}(\theta)$ and $T_{+}(-\theta)$. We parametrize
$T_{+}$ such a way that for its physical domain ($\theta<0$) its
argument is always positive. Transmission factors satisfy unitarity
and defect crossing symmetry \cite{DefBound}\begin{equation}
T_{+}(-\theta)=T_{-}^{-1}(\theta)\quad;\qquad T_{-}(\theta)=T_{+}(i\pi-\theta)\label{Defprop}\end{equation}
The multiparticle transition amplitude can be derived by introducing
the defect operator $D^{+}$ and the following relations in the ZF
algebra: \[
A^{+}(\theta)D^{+}=T_{-}(\theta)D^{+}B^{+}(\theta)\qquad;\quad D^{+}B^{+}(-\theta)=T_{+}(\theta)A^{+}(-\theta)D^{+}\]
A defect is parity symmetric if $T_{-}(\theta)=T_{+}(\theta)$. Clearly
$A^{+}(\theta=0)$ satisfies the properties of $D^{+}$ with $T_{-}(\theta)=S(\theta)=T_{+}(\theta)$
. Thus a standing particle can be considered as the prototype of a
parity symmetric defect.

\subsection{Coordinate dependence of the form factors}

The form factor of a local operator $\mathcal{O}(x,t)$ is its matrix
element between asymptotic states:\[
\langle\theta_{m^{'}}^{'},\dots,\theta_{n^{'}+1}^{'};\theta_{n^{'}}^{'},\dots,\theta_{1}^{'}\vert\mathcal{O}(x,t)\vert\theta_{1},\dots,\theta_{n};\theta_{n+1},\dots,\theta_{m}\rangle\]
where the adjoint state is defined to be \[
\langle\theta_{m^{'}}^{'},\dots,\theta_{n^{'}+1}^{'};\theta_{n^{'}}^{'},\dots,\theta_{1}^{'}\vert=\langle0\vert B(\theta_{m^{'}}^{'})\dots B(\theta_{n^{'}+1}^{'})DA(\theta_{n^{'}}^{'})\dots A(\theta_{1}^{'})\]
Strictly speaking the form factor is defined only for initial/final
states (i.e. for decreasingly/increasingly ordered arguments) but
using the ZF algebra we can generalize them for any values and orders
of the rapidities. 

The multiparticle asymptotic states are eigenstates of the conserved
energy. This fact can be formulated in the language of the ZF algebra
as \[
[H,A^{+}(\theta)]=m\cosh\theta\, A^{+}(\theta)\qquad;\qquad[H,D^{+}]=e_{D}D^{+}\]
In the second equation we supposed that the vacuum containing the
defect has energy $e_{D}$. Classical considerations together with
the topological nature of the defect suggest the existence of a conserved
momentum with properties\[
[P,A^{+}(\theta)]=m\sinh\theta\, A^{+}(\theta)\qquad;\qquad[P,D^{+}]=p_{D}D^{+}\]
Thus, opposed to a general boundary theory, the defect breaks translation
invariance by having a nonzero momentum eigenvalue $p_{D}$ and not
by destroying the existence of the momentum itself. As a consequence
the time and space dependence of the form factor can be obtained as\begin{eqnarray*}
\langle\theta_{m^{\prime}}^{\prime},\dots,\theta_{n^{\prime}+1}^{\prime};\theta_{n^{\prime}}^{\prime},\dots,\theta_{1}^{\prime}\vert\mathcal{O}(x,t)\vert\theta_{1},\dots,\theta_{n};\theta_{n+1},\dots,\theta_{m}\rangle=\\
e^{it\Delta E-ix\Delta P}F_{(n^{\prime},m^{\prime})(n,m)}^{\mathcal{O}}(\theta_{n\prime+m\prime}^{\prime},...,\theta_{n\prime+1}^{\prime};\theta_{n\prime}^{\prime},...,\theta_{1}^{\prime}\vert\theta_{1},\dots,\theta_{n};\theta_{n+1},...,\theta_{n+m})\end{eqnarray*}
 where $\Delta E=m(\sum_{j}\cosh\theta_{j}-\sum_{j^{\prime}}\cosh\theta_{j^{\prime}}^{\prime})$
and $\Delta P=m(\sum_{j}\sinh\theta_{j}-\sum_{j^{\prime}}\sinh\theta_{j^{\prime}}^{\prime})$.
The very same simple space and time dependence can be seen also in
a theory without the defect and it is substantially different from
what we would expect from a general boundary theory where even the
one point function has a nontrivial space-dependence. These considerations
remain valid for operators inserted at the defect $\mathcal{O}(t)=\mathcal{O}(0,t)$,
too.

\subsection{Crossing transformation of defect form factors}

The properties and analytical structure of the form factor $F_{(n^{\prime}m^{\prime}),(nm)}$
can be derived via the reduction formula from the correlation functions
similarly to the boundary case \cite{BFF}. Instead of going to the
details of the calculation of \cite{BFF} we note that all equations
follow from the defining relations of the ZF algebra and the locality
of the operator $[\mathcal{O}(0,0),A^{+}(\theta)]=0$ except the crossing
relation. This missing relation reads as \begin{eqnarray*}
F_{(n\prime,m\prime)(n,m)}^{\mathcal{O}}(\theta_{n\prime+m\prime}^{\prime},...,\theta_{n\prime+1}^{\prime};\theta_{n\prime}^{\prime},...,\theta_{1}^{\prime}\vert\theta_{1},\dots,\theta_{n};\theta_{n+1},...,\theta_{n+m})=\\
F_{(n\prime,m\prime+1)(n,m-1)}^{\mathcal{O}}(\theta_{n+m}+i\pi,\theta_{n\prime+m\prime}^{\prime},...,\theta_{n\prime+1}^{\prime};\theta_{n\prime}^{\prime},...,\theta_{1}^{\prime}\vert\theta_{1},\dots,\theta_{n};\theta_{n+1},...,\theta_{n+m-1})\end{eqnarray*}
and can be obtained as follows: We fold the system \cite{DefBound}
to a boundary one: $B^{+}(\theta)\leftrightarrow\tilde{B}^{+}(-\theta)$,
and consider $A^{+}$ and $\tilde{B}^{+}$ as creation operators of
two different type of particles which scatter trivially on each other.
Now we apply the crossing equation of $\tilde{B}^{+}$ for the resulting
boundary form factor \cite{BFF}. If we fold back the system to the
original defect theory we obtain the defect crossing equation above. 

By analyzing the crossing equation of the particle $A^{+}$ instead
of $B^{+}$ we obtain \begin{eqnarray*}
F_{(n\prime,m\prime)(n,m)}^{\mathcal{O}}(\theta_{n\prime+m\prime}^{\prime},...,\theta_{n\prime+1}^{\prime};\theta_{n\prime}^{\prime},...,\theta_{1}^{\prime}|\theta_{1},\dots,\theta_{n};\theta_{n+1},...,\theta_{n+m}) & =\\
F_{(n\prime+1,m\prime)(n-1,m)}^{\mathcal{O}}(\theta_{n\prime+m\prime}^{\prime},...,\theta_{n\prime+1}^{\prime};\theta_{n\prime}^{\prime},...,\theta_{1}^{\prime},\theta_{1}-i\pi|\theta_{2},\dots,\theta_{n};\theta_{n+1},...,\theta_{n+m})\end{eqnarray*}
This crossing equation can also be obtained from (\ref{eq:crossing}).
Using any of the crossing equations above we can express all form
factors in terms of the one-sided form factors: \[
F_{(n,m)}^{\mathcal{O}}(\theta_{1},\dots,\theta_{n};\theta_{n+1},...,\theta_{n+m}):=F_{(0,0)(n,m)}^{\mathcal{O}}(;\vert\theta_{1},\dots,\theta_{n};\theta_{n+1},...,\theta_{n+m})\]
 on which we focus in the rest of the paper. The properties of this
form factor follows from the ZF algebra relations and from the crossing
relations and we postulate them in the next subsection as axioms.

\subsection{Defect form factor axioms}

The matrix elements of local operators satisfy the following axioms:

\bigskip

I. Transmission:

\begin{center}
\[
F_{(n,m)}^{\mathcal{O}}(\theta_{1},\dots,\theta_{n};\theta_{n+1},...,\theta_{n+m})=T_{-}(\theta_{n})F_{(n-1,m+1)}^{\mathcal{O}}(\theta_{1},\dots,\theta_{n-1};\theta_{n},\theta_{n+1},...,\theta_{n+m})\]
\includegraphics[height=3cm]{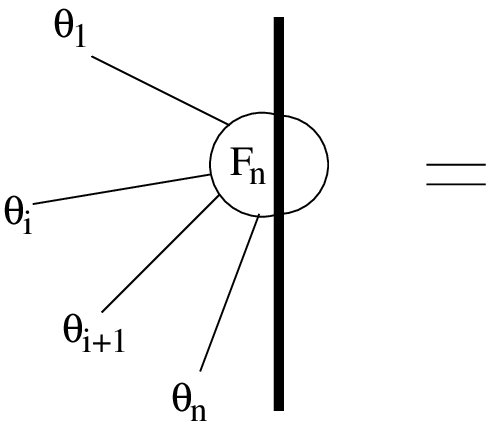}~~~~~~~\includegraphics[height=3cm]{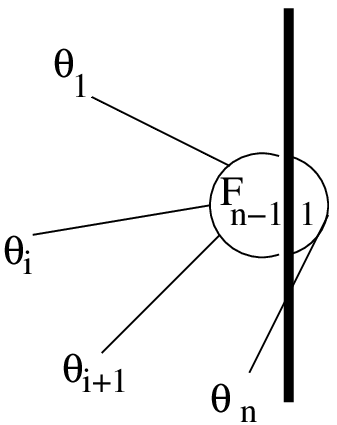}
\par\end{center}

By means of this axiom we can express every form factor in terms of
the \emph{elementary} one\[
F_{n}^{\mathcal{O}}(\theta_{1},\dots,\theta_{n})=F_{(n,0)}^{\mathcal{O}}(\theta_{1},\dots,\theta_{n};)\]
It satisfies the further axioms:

\bigskip

II. Permutation:\[
F_{n}^{\mathcal{O}}(\theta_{1},\dots\theta_{i},\theta_{i+1},\dots,\theta_{n})=S(\theta_{i}-\theta_{i+1})F_{n}^{\mathcal{O}}(\theta_{1},\dots\theta_{i+1},\theta_{i},\dots,\theta_{n})\]

\begin{center}
\includegraphics[height=3cm]{bd1}~~~~~~~~~\includegraphics[height=3cm]{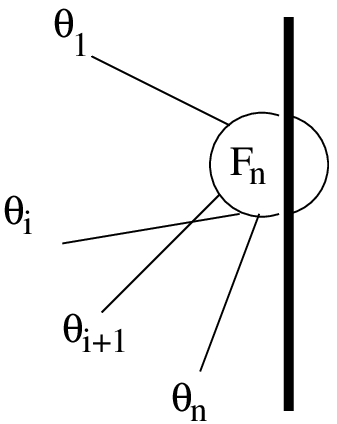}
\par\end{center}

\bigskip

III. Periodicity:\[
F_{n}^{\mathcal{O}}(\theta_{1},\theta_{2},\dots,\theta_{n})=F_{n}^{\mathcal{O}}(\theta_{2},\dots\theta_{n},\dots,\theta_{1}-2i\pi)\]

\begin{center}
\includegraphics[height=3cm]{bd1}~~~~~~~~\includegraphics[height=3cm]{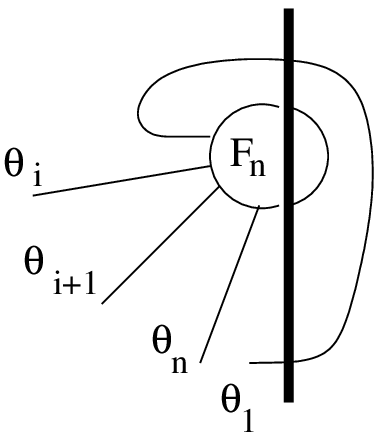}
\par\end{center}

The physical singularities can be formulated as follows. 

\bigskip

IV. Kinematical singularity:\[
-i\mbox{Res}_{\theta=\theta\prime}F_{n+2}^{\mathcal{O}}(\theta+i\pi,\theta^{\prime},\theta_{1},...,\theta_{n})=\left(1-\prod_{j=1}^{n}S(\theta-\theta_{j})\right)F_{n}^{\mathcal{O}}(\theta_{1},...,\theta_{n})\]

\begin{center}
\includegraphics[height=3cm]{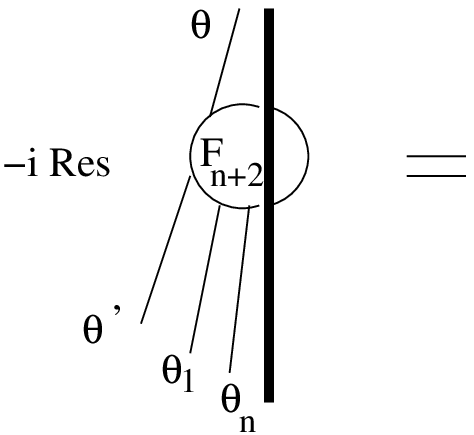}~~~\includegraphics[height=3cm]{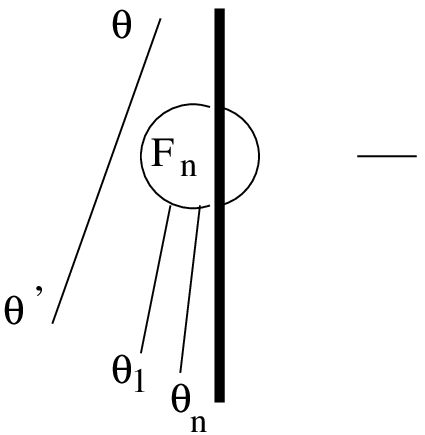}~~~\includegraphics[height=3cm]{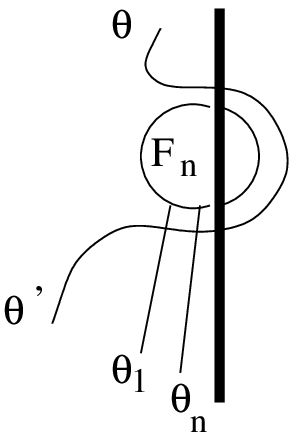}
\par\end{center}

\bigskip

V. Dynamical bulk singularity:\[
-i\mbox{Res}_{\theta^{\prime}=\theta}F_{n+2}^{\mathcal{O}}(\theta^{\prime}+\frac{i\pi}{3},\theta-\frac{i\pi}{3},\theta_{1},\ldots,\theta_{n})=\Gamma F_{n+1}^{\mathcal{O}}(\theta,\theta_{1},\ldots,\theta_{n})\]

\begin{center}
\includegraphics[height=3cm]{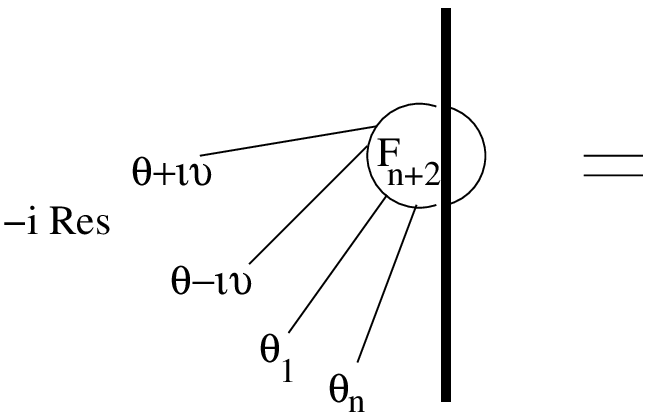}~~~~~~\includegraphics[height=3cm]{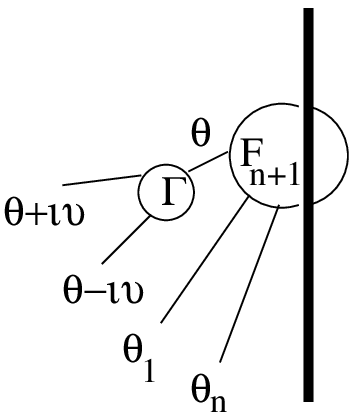}
\par\end{center}

where $\Gamma$ is the 3 particle on-shell coupling.

\bigskip

VI. Dynamical defect singularity:\[
-i\mbox{Res}_{\theta=iu}F_{n+1}^{\mathcal{O}}(\theta_{1},\ldots,\theta_{n},\theta)=ig\tilde{F}_{n}^{\mathcal{O}}(\theta_{1},\ldots,\theta_{n})\]

\begin{center}
\includegraphics[height=3cm]{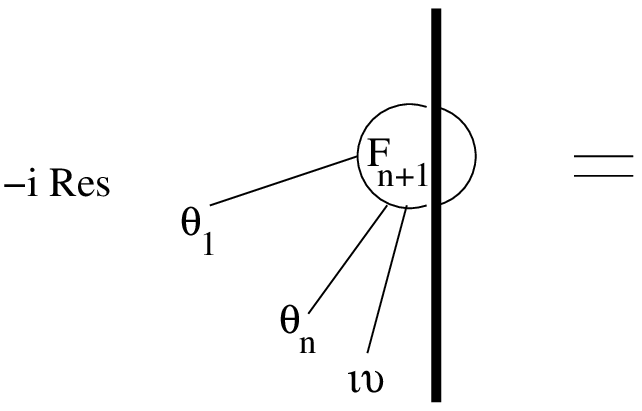}~~~~~~~\includegraphics[height=3cm]{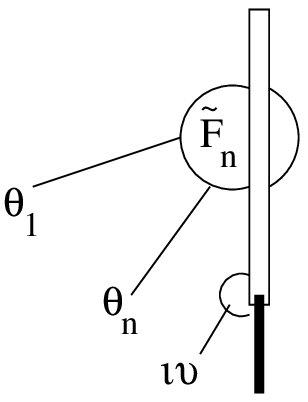}
\par\end{center}

where $g$ is the defect bound-state coupling.

A few remarks are in order: Although the form factor axioms (II-V)
are the same as the axioms of the form factors in a theory without
the defect \cite{FFbook}, the axioms (I,VI) are different and in
general defect theories will have different solutions. An exception
is the invisible defect $T_{-}(\theta)=1$ when we recover the usual
form factor equation providing a consistency check for our axioms.
Another consistency check can be obtained by considering a standing
particle as the defect. Then $T_{\pm}(\theta)=S(\theta)$ and the
two additional axioms become part of the old ones: (I,VI) will be
special cases of (II,V), respectively.

\section{Form factor solutions, two point functions}

In this section we determine the solutions of the form factor equations
for operators localized in the bulk and at the defect. For operators
localized in the bulk the solutions can be built up form the bulk
form factors and from the transmission factors. Using these form factor
solutions we determine the spectral representation of the two point
function for the situations when the operators are localized on the
same or on the opposite sides of the defect. Finally, for operators
localized on the defect we outline the strategy for the general solution.

\subsection{Bulk operators}

The form factor axioms for $F_{n}$ are the same as in the bulk so
we expect to use the bulk form factor solutions. Clearly we have to
make a distinction whether the operator are localized on the left,
or on the right of the defect. If the operator is localized on the
left then particles arriving from the left can reach the operator
without crossing the defect. Since the defect is topological we can
change its location without altering the form factor (as far as we
do not cross the insertion point of the operator). Shifting then the
defect far away we expect to obtain the form factors of the bulk theory. 

\begin{center}
\includegraphics[height=3cm]{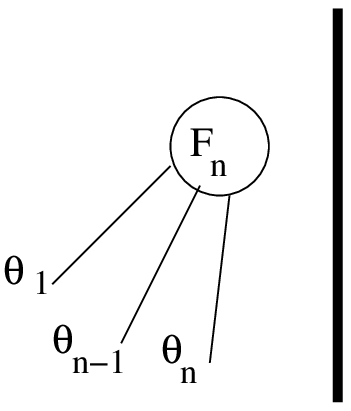}~~~~~\includegraphics[height=3cm]{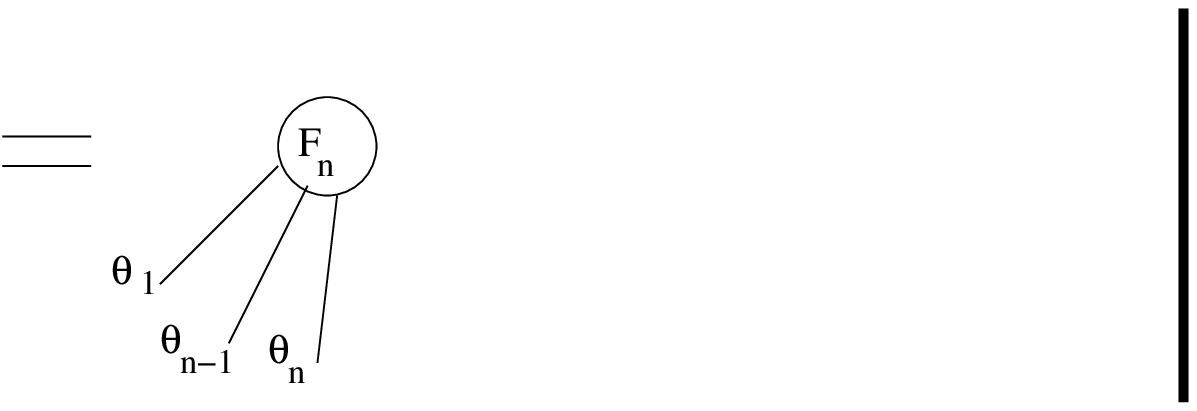}
\par\end{center}

Thus we can conlude that for the initial state $\theta_{1}>\dots>\theta_{n}>0$
the defect form factor coincides with the bulk form factor. Let us
denote the solutions of the bulk form factor equations by $B_{n}(\theta_{1},\dots,\theta_{n})$.
Then we claim that for an operator localized on the left ($\mathcal{O}^{<}$
for short) we have \begin{equation}
F_{n}^{<}(\theta_{1},\dots,\theta_{n})=B_{n}(\theta_{1},\dots,\theta_{n})=F_{(n,0)}^{<}(\theta_{1},\dots,\theta_{n};)\label{eq:bulkleft}\end{equation}
By using the transmission axiom and the crossing relation we can express
all other matrix elements in terms of the bulk matrix element and
the transmission factor. If the operator is localized on the right
of the defect ($\mathcal{O}^{>}$) then, by similar argumentations,
we expect the defect form factor to coincide with the bulk form factor
for particles coming from the right. Those initial states have the
ordering $0>\theta_{1}>\dots>\theta_{n}$ and the form factor is then
\[
F_{(0,n)}^{>}(;\theta_{1},\dots,\theta_{n})=B_{n}(\theta_{1},\dots,\theta_{n})\]
The solution for the elementary defect form factor for operators localized
on the right thus turns out to be \begin{equation}
F_{n}^{>}(\theta_{1},\dots,\theta_{n})=\prod_{i}T_{-}(\theta_{i})B_{n}(\theta_{1},\dots,\theta_{n})\label{eq:bulkright}\end{equation}
which satisfies all the bulk form factor axioms but \emph{does not
coincide} with the bulk form factor solution.

Having calculated the form factor solutions we use them to construct
the two point functions, which, for operators localized on the opposite
side of the defect, will be intrinsically different from the one without
the defect. We analyze the following two point function \[
\langle;\vert\mathcal{O}_{1}(x_{1},t_{1})\mathcal{O}_{2}(x_{2},t_{2})\vert;\rangle\]
where we denote by $\vert;\rangle$ the vacuum of the defect theory.
Formally $\vert;\rangle=D^{+}\vert0\rangle$. Now we insert the resolution
of the identity. It can be composed both from initial and from final
states and for definiteness we choose initial states. It is instructive
to list the possible states. If we have no particles we have only
the vacuum: $\vert;\rangle$. One particle states can be of two types,
depending on whether the particle arrives from the left or from the
right: $\vert\theta;\rangle$ for $\theta>0$ and $\vert;\theta\rangle$
for $\theta<0$. A general $N=n+m$ particle state $\vert\theta_{1},\dots,\theta_{n};\theta_{n+1},\dots,\theta_{n+m}\rangle$
with $\theta_{1}>\dots>\theta_{n}>0>\theta_{n+1}>\dots>\theta_{m}$
has to cover all possible cases ranging from $n=0$ to $n=N$ . The
two point function then can be written formally as \[
\langle;\vert\mathcal{O}_{1}(x_{1},t_{1})\mathcal{O}_{2}(x_{2},t_{2})\vert;\rangle=\sum_{N=0}^{\infty}\langle;\vert\mathcal{O}_{1}(0,0)\vert N\rangle\langle;\vert\mathcal{O}_{2}^{*}(0,0)\vert N\rangle^{*}e^{iE(N)(t_{1}-t_{2})-iP(N)(x_{1}-x_{2})}\]
We have to specify the integration ranges for the multiparticle state
$N$. Originally we have to integrate only for the multiparticle momentum
range of the \emph{initial }states. If we exchange the order from
$\theta_{1}>\theta_{2}$ to the nonphysical $\theta_{2}>\theta_{1}$then
the form factor of $\mathcal{O}_{1}$ picks up a factor $S(\theta_{1}-\theta_{2})$
while that of $\mathcal{O}^{*}$ the inverse factor $S^{*}(\theta_{1}-\theta_{2})$,
so the integrand is a symmetric function. For each integration with
$\theta_{1}>0$ we also have an analogous integration for $\theta_{1}<0$
. Their contributions differ by a factor $T_{+}(\theta_{1})$ for
the form factor of $\mathcal{O}_{1}$ and by the inverse $T_{+}^{*}(\theta_{1})$
for $\mathcal{O}_{2}$. As a consequence we can express the correlator
in terms of the elementary form factors $F_{n}$ as: \begin{eqnarray*}
\langle;\vert\mathcal{O}_{1}(x_{1},t_{1})\mathcal{O}_{2}(x_{2},t_{2})\vert;\rangle=\\
 &  & \hspace{-3cm}\sum_{n=0}^{\infty}\frac{1}{n!}\int_{-\infty}^{\infty}\frac{d\theta_{1}}{2\pi}\dots\int_{-\infty}^{\infty}\frac{d\theta_{n}}{2\pi}F_{n}^{O_{1}}(\theta_{1},\dots,\theta_{n})F_{n}^{O_{2}^{*}}(\theta_{1},\dots,\theta_{n})^{*}e^{iE(n)(t_{1}-t_{2})-iP(n)(x_{1}-x_{2})}\end{eqnarray*}
Although we transported the operators $\mathcal{O}_{1}$ and $\mathcal{O}_{2}$
into the origin, the form factor solutions remember whether the operators
are localized on the left or on the right of the defect. 

If both operators are localized on the left, ($x_{1}<0,\, x_{2}<0$)
then the elementary form factors are the same as the bulk form factors
(\ref{eq:bulkleft}) and we can conclude that the two point function
is\emph{ exactly the same} as the bulk two point function \begin{eqnarray*}
\langle;\vert\mathcal{O}_{1}(x_{1},t_{1})\mathcal{O}_{2}(x_{2},t_{2})\vert;\rangle=\\
 &  & \hspace{-3cm}\sum_{n=0}^{\infty}\frac{1}{n!}\int_{-\infty}^{\infty}\frac{d\theta_{1}}{2\pi}\dots\int_{-\infty}^{\infty}\frac{d\theta_{n}}{2\pi}B_{n}^{O_{1}}(\theta_{1},\dots,\theta_{n})B_{n}^{O_{2}^{*}}(\theta_{1},\dots,\theta_{n})^{*}e^{iE(n)(t_{1}-t_{2})-iP(n)(x_{1}-x_{2})}\end{eqnarray*}
This is intuitively clear: we can transport the defect to infinity
without crossing any of the insertion points thus leaving invariant
the two point function. When the defect is at infinity it does not
influences the two point function which then has to be the same as
in the bulk. The same result can be obtained when both operators are
localized on the right of the defect. 

If the operators are localized on different sides of the defect ($x_{1}<0$,
$x_{2}>0$) then additinally to (\ref{eq:bulkleft}) we also have
to use (\ref{eq:bulkright}). As a result the two point function is
expressed in terms of the bulk form factors $B_{n}$ and the transmission
matrix $T_{-}(\theta)$ as \begin{eqnarray*}
\langle;\vert\mathcal{O}_{1}(x_{1},t_{1})\mathcal{O}_{2}(x_{2},t_{2})\vert;\rangle=\\
 &  & \hspace{-5cm}\sum_{n=0}^{\infty}\frac{1}{n!}\int_{-\infty}^{\infty}\frac{d\theta_{1}}{2\pi}T_{-}(\theta_{1})\dots\int_{-\infty}^{\infty}\frac{d\theta_{n}}{2\pi}T_{-}(\theta_{n})B_{n}^{O_{1}}(\theta_{1},\dots,\theta_{n})B_{n}^{O_{2}^{*}}(\theta_{1},\dots,\theta_{n})^{*}e^{iE(n)(t_{1}-t_{2})-iP(n)(x_{1}-x_{2})}\end{eqnarray*}
This is the main result of this section. This formula shows how the
correlation function can be calculated in the presence of an integrable
defect in terms of the transmission factor and the bulk form factors.
It can be generalized to any correlators localized in the bulk using
the resolution of the identity together with the exact form factors
(\ref{eq:bulkleft}) and (\ref{eq:bulkright}). It cannot be applied,
however, for operators localized on the defect, which is the subject
of the next subsection.

\subsection{Defect operators}

We have seen that although the minimal form factors $F_{n}$ are subject
to the same requirement as the bulk form factors they are not necessarily
the same. In this subsection we develop a general methodology to determine
the defect form factors. Let us analyze them for increasing particle
numbers:

The first form factor is the vacuum expectation value of a defect
field \[
\langle;\vert\mathcal{O}(t)\vert;\rangle=F_{0}\]

The one particle form factor is defined to be \[
\langle;\vert\mathcal{O}(t)\vert\theta;\rangle=F_{1}(\theta)\quad;\qquad\langle;\vert\mathcal{O}(t)\vert;\theta\rangle=T_{-}(\theta)F_{1}(\theta)\]
 Contrary to the bulk case it has a nontrivial rapidity dependence:
it is not natural to take $F_{1}$ to be constant. In a parity invariant
theory for a parity symmetric operators, for example, we have $F_{1}(\theta)=T_{-}(-\theta)F_{1}(-\theta)$.
If parity is broken then $F_{1}(\theta)$ can be an arbitrary defect
condition dependent $2\pi i$ periodic function. The only restriction
came from the defect bound-state axiom (VI): it must have a pole at
$iv$ whenever $T_{-}(\theta)$ has a pole correspondig to a bound-state.
Let us denote the minimal function which satisfies this requirement
by $d(\theta)$. The general form of the one particle form factor
is then \[
\langle;\vert\mathcal{O}(t)\vert\theta;\rangle=d(\theta)Q_{1}^{\mathcal{O}}(u)\quad;\qquad u=e^{\theta}\]
where $d(\theta)$ depends on the defect condition, while $Q_{1}(u)$
depends on the operator we are dealing with. 

The two particle form factor must also have a singularity at $i\nu$
and additionally it satisfies the bulk form factor axioms so we expect
it to be written into the form \[
F_{2}(\theta_{1},\theta_{2})=d(\theta_{1})d(\theta_{2})f_{min}(\theta_{1}-\theta_{2})\frac{Q_{2}^{\mathcal{O}}(u_{1},u_{2})}{u_{1}+u_{2}}\]
where $f_{min}(\theta)$ is the minimal solution of the bulk two particle
form factor equations\[
f_{min}(\theta)=S(\theta)f_{min}(-\theta)\quad;\qquad f_{min}(i\pi-\theta)=f_{min}(i\pi+\theta)\]
Here $Q_{2}$ is a symmetric function in $u_{i}=e^{\theta_{i}}$,
containing a factor $(u_{1}+u_{2})$ to kill the denominator, what
we introduced to conform with the kinematical singularity appearing
at higher levels only. Taking into account the general paramterization
of the bulk and boundary form factors together with the dynamical
and kinematical singularity axioms we parametrize our minimal defect
form factors as \begin{equation}
F_{n}(\theta_{1},\dots,\theta_{n})=\prod_{i}d(\theta_{i})\prod_{i<j}\frac{f_{min}(\theta_{i}-\theta_{j})}{u_{i}+u_{j}}Q_{n}(u_{1},\dots,u_{n})\label{eq:dffansatz}\end{equation}
where $Q_{n}(u_{1},\dots,u_{n})$ is a symmetric function expected
to be a polynomial, if there is no bulk dynamical singularity. If
there is such a singularity we have to include the corresponding singularity
into $f_{min}$ . The dependence on the defect condition is contained
in $d(\theta)$ with possible defect bound-state singularities, while
the dependence on the operator in $Q$. If for instance the defect
is the invisible defect with $T_{\pm}=1$ then $d=1$ and we recover
the solution of the bulk form factor equation as it should be. From
the kinematical singularity equations recursion relations can be obtained
among the polynomials $Q_{n+2}$ and $Q_{n}$.

\section{Model studies}

In this section we analyze the solutions of the defect form factor
axioms for the free boson and for the Lee-Yang models.

\subsection{Free boson }

The purely transmitting free bosonic theory was analyzed in \cite{BS}
as the limiting case of the sinh-Gordon theory. Let us recall its
solution. The Lagrangian of the model reads as 

\begin{eqnarray*}
\mathcal{L} & = & \Theta(-x)\left[\frac{1}{2}(\partial_{\mu}\Phi_{-})^{2}-\frac{m^{2}}{2}\Phi_{-}^{2}\right]+\Theta(x)\left[\frac{1}{2}(\partial_{\mu}\Phi_{+})^{2}-\frac{m^{2}}{2}\Phi_{+}^{2}\right]\\
 &  & -\frac{\delta(x)}{2}\left(\Phi_{+}\dot{\Phi}_{-}-\Phi_{-}\dot{\Phi}_{+}+m\left[(\cosh\mu)\left(\Phi_{+}^{2}+\Phi_{-}^{2}\right)+2(\sinh\mu)\,\Phi_{+}\Phi_{-}\right]\right)\end{eqnarray*}
where $\Phi_{\pm}$ are the fields living on the right/left part of
the defect and $\mu$ is a free parameter. By varying the action we
obtain the free equation of motion in the bulk \begin{equation}
(\partial_{x}^{2}-\partial_{t}^{2})\Phi_{\pm}=m^{2}\Phi_{\pm}\label{eq:bulkeom}\end{equation}
and the defect condtions: \begin{eqnarray}
\pm\partial_{t}\Phi_{\pm}\mp\partial_{x}\Phi_{\mp} & = & m(\sinh\mu)\,\Phi_{\pm}+m(\cosh\mu)\,\Phi_{\mp}\label{eq:defcond}\end{eqnarray}
Since $\Phi_{\pm}$ are free fields they have an expansion in terms
of plane waves and creation/annihilation operators \[
\Phi_{\pm}(x,t)=\int_{-\infty}^{\infty}\frac{dk}{2\pi}\frac{1}{2\omega(k)}\left(a_{\pm}(k)e^{ikx-i\omega(k)t}+a_{\pm}^{+}(k)e^{-ikx+i\omega(k)t}\right)\quad;\quad\omega(k)=\sqrt{k^{2}+m^{2}}\]
 where the $a,a^{+}$ operators are adjoint of each other with commutators:\[
[a_{\pm}(k),a_{\pm}^{+}(k^{'})]=2\pi2\omega(k)\delta(k-k^{'})\]
They are not independent, the defect condition connects them as \[
a_{\pm}(\pm k)=T_{\mp}(k)a_{\mp}(\pm k)\quad;\qquad T_{\mp}(k)=-\frac{m\sinh\mu\mp i\omega(k)}{m\cosh\mu-ik}\quad;\quad k>0\]
This shows that the defect is purely transmitting, that is we do not
have any reflected wave. The transmission factor in the rapidity parametrization
($k=m_{\mathrm{cl}}\sinh\theta)$ can be written also in the following
form: \[
T_{-}(\theta)=-i\frac{\sinh(\frac{\theta}{2}-\frac{i\pi}{4}+\frac{\mu}{2})}{\sinh(\frac{\theta}{2}+\frac{i\pi}{4}+\frac{\mu}{2})}=\frac{1+wu}{1-wu}\]
where $w=ie^{\mu}$ and $u=e^{\theta}$. Sometimes we also use $\bar{w}=w^{-1}$
and $\bar{u}=u^{-1}$. In the next subsection we will set $m=1$ and
use dimensionless quantities.

\subsubsection{Form factors }

In the free boson model, we have the advantage that we can explicitly
calculate the form factors of all the operators, and then check that
they satisfy the defect form factor axioms. Additionally we can also
confirm that we have as many polynomial solutions of the axioms as
many local operators exist in the theory. 

We work with the Euclidien version of the theory ($t=iy$) and introduce
complex coordinates $2z=y+ix$ , $2\bar{z}=y-ix$ . Using the explicit
expressions of $\Phi_{\pm}(z,\bar{z})$ we are able to calculate the
form factors. We analyze them for increasing particle number and compare
to the form factor solution. 

The one particle form factors turn out to be: 

\[
F_{1}^{\Phi_{-}}=\langle0|\Phi_{-}(z,\bar{z})|a_{-}(\theta)\rangle=e^{zu+\bar{z}\bar{u}}\]
\[
F_{1}^{\Phi_{+}}=\langle0|\Phi_{+}(z,\bar{z})|a_{-}(\theta)\rangle=e^{zu+\bar{z}\bar{u}}T_{-}(\theta)\]

from which it is easy to calculate the defect form factors of the
derivative of the elementary fields:

\[
\langle0|\partial^{n}\Phi_{-}(0)|a_{-}(\theta)\rangle=u^{n}\quad;\qquad\langle0|\bar{\partial}^{n}\Phi_{-}(0)|a_{-}(\theta)\rangle=\bar{u}^{n}\]
\[
\langle0|\partial^{n}\Phi_{+}(0)|a_{-}(\theta)\rangle=u^{n}T_{-}(\theta)\quad;\qquad\langle0|\bar{\partial}^{n}\Phi_{+}(0)|a_{-}(\theta)\rangle=\bar{u}^{n}T_{-}(\theta)\]
where $\partial=\partial_{z}$ and $\bar{\partial}=\partial_{\bar{z}}$.
We can unify this notation by $\partial^{-n}=\bar{\partial}^{n}$.
It is instructive to see how we can recover these form factors from
the solution of the form factor axioms. Using the parametrization
of the form factors in terms of $d(\theta)$ see equation (\ref{eq:dffansatz}),
we know that at level 1 the solutions of the form factor axioms have
the form:

\[
F_{1}(\theta)=d(\theta).Q_{1}(\theta)\]
Thus if we choose\[
d(\theta)=\frac{1}{1-wu}\]
 we obtain

\[
Q_{1}^{\partial^{n}\Phi_{-}}(\theta)=u^{n}(1-wu)\quad;\qquad Q_{1}^{\partial^{n}\Phi_{+}}(\theta)=u^{n}(1+wu)\]
Naively it seems we have less polynomial solutions of the form factor
equations than operators. We have extra relations among the form factors
and they originate from \[
\partial\bar{\partial}\Phi_{\pm}=\Phi_{\pm}\quad;\qquad\bar{\partial}\Phi_{+}-\bar{\partial}\Phi_{-}=w(\Phi_{+}+\Phi_{-})\quad;\qquad\partial\Phi_{-}+\partial\Phi_{+}=\bar{w}(\Phi_{+}-\Phi_{-})\]

However, these relations are satisfied due to the bulk equation of
motion (\ref{eq:bulkeom}) and the defect conditions (\ref{eq:defcond}).
Note that the form factor solutions are even more simple in terms
of $\phi=\Phi_{+}+\Phi_{-}$ and $\bar{\phi}=\Phi_{+}-\Phi_{-}$.
Actually $\bar{\phi}$ is not idependent since $\bar{\phi}=2w\partial\phi$.
Their form factors read as: \[
Q_{1}^{\partial^{n}\phi}=u^{n}\qquad;Q_{1}^{\partial^{n}\bar{\phi}}=wu^{n+1}\]
Observe also that by changing the sign of the exponential coupling
$e^{\mu}\to-e^{\mu}$ the left right fields are interchanged $\Phi_{\pm}\leftrightarrow\Phi_{\mp}$,
as follows from the discrete symmetries of the Lagrangian. 

In the form factor boostrap the general parametrization without kinematical
singularity is \[
F_{n}(\theta_{1},\dots,\theta_{N})=\prod_{i=1}^{N}d(\theta_{i})Q_{N}(x_{1},\dots,x_{N})\]
 Thus we can read off the corresponding form factor solution directly\[
Q_{N}=u_{1}^{n_{1}}\dots u_{N}^{n_{N}}(1-wu_{1})\dots(1-wu_{k})(1+wu_{k+1})\dots(1+wu_{N})+\mbox{permutations}\]
Since the scattering matrix in the free boson theory is trivial $S=1$,
the form factors of different levels are not connnected to each other,
and in this way we solved the theory completely. 

In terms of the field $\phi$ the form factor solutions are exactly
the same as in the bulk free bosonic theory: \[
Q_{N}=u_{1}^{n_{1}}\dots u_{N}^{n_{N}}+\mbox{permutations}\]
thus we obtain exactly the same number of polynomial solution of the
form factor axioms as many independent local operator exists in the
theory.

\subsection{Defect scaling Lee-Yang model}

The scaling Lee-Yang model can be defined as a perturbation of the
$\mathcal{M}_{(2,5)}$ conformal minimal model with central charge
$c=-\frac{22}{5}$. It contains two chiral representations of the
Virasoro algebra, $V_{0},V_{1}$ with highest weights $0$ and $-\frac{1}{5}$,
respectively. The fusion rules can be summarized as: $N_{0i}^{i}=N_{i0}^{i}=1$
and $N_{11}^{i}=1$ for $i=0,1$ and all others are zero. The Hilbert
space on the torus corresponds to the (diagonal) modular invariant
partition function and contains modules corresponding to the $Id$
and the $\Phi(z,\bar{z})$ primary fields with weights $(0,0)$ and
$(-\frac{1}{5},-\frac{1}{5})$:\begin{equation}
\mathcal{H}=V_{0}\otimes\bar{V}_{0}+V_{1}\otimes\bar{V}_{1}\label{eq:bulkH}\end{equation}
 The only relevant perturbation by the field $\Phi$ results in the
simplest scattering theory with one neutral particle of mass $m$
and scattering matrix \cite{SYLbulk} \[
S(\theta)=\frac{\sinh\theta+i\sin\frac{\pi}{3}}{\sinh\theta-i\sin\frac{\pi}{3}}\]
The pole at $\theta=\frac{i\pi}{3}$ (with residue $\Gamma^{2}$)
shows that the particle can form a bound-state. The relation\[
S(\theta+i\frac{\pi}{3})S(\theta-i\frac{\pi}{3})=S(\theta)\]
however, implies that the bound-state is the original particle itself
and the bulk bootstrap is closed.

\subsection{Integrable defects}

Two types of topological defects can be introduced in the $\mathcal{M}_{(2,5)}$
minimal model \cite{PZ,QRW}. They can be considered as operators
acting on the bulk Hilbert space (\ref{eq:bulkH}) commuting with
the action of the left and right Virasoro generators. They have to
act diagonally on each factor in (\ref{eq:bulkH}) and satisfy a Cardy
type condition. This leads to two choices which can be labelled by
the same way as the bulk fields: $(0,0)$ and $(1,1)$. After making
a modular transformation the defect is inserted in space. The operators
living on the defect can be described as \cite{Ingo} \[
\mathcal{H}^{(a,a)}=\sum_{i,j}(V_{i}\otimes\bar{V}_{j})^{\oplus(\sum_{c\in\{0,1\}}N_{ia}^{c}N_{cj}^{a})}\]
For the topological defect labeled by $(0,0)$ the operator space
turns out to be \[
\mathcal{H}^{(0.0)}=V_{0}\otimes\bar{V}_{0}+V_{1}\otimes\bar{V}_{1}\]
and coincides with the bulk Hilbert space. This defect is the trivial
(invisible) defect. 

For the other defect labeled by $(1,1)$ we obtain \[
\mathcal{H}^{(1,1)}=V_{0}\otimes\bar{V}_{0}+V_{1}\otimes\bar{V}_{0}+V_{0}\otimes\bar{V}_{1}+2\, V_{1}\otimes\bar{V}_{1}\]
For each of the representation spaces we associate a primary field
$Id,\,\varphi(z),\bar{\,\varphi}(\bar{z}),\,\Phi_{-}(z,\bar{z}),\,\Phi_{+}(z,\bar{z})$
with highest weights $(0,0),(-1/5,0),(0,-1/5),(-1/5,-1/5),(-1/5,-1/5)$,
respectively. The nonchiral fields $\Phi_{\pm}(z,\bar{z})$ can be
considered as the left/right limits of the bulk field $\Phi(z,\bar{z})$
on the defect. 

The bulk perturbation by $\Phi$ in the defect conformal field theory
does not break integrability. In the case of the trivial defect the
transmission factor is simply the identity $T=1.$ In the case of
the defect labeled by $(1,1)$ we can introduce a one parameter family
of defect perturbations as well by properly harmonizing the coefficients
of the $\varphi(z)$, $\bar{\varphi}(\bar{z})$ and $\Phi(z,\bar{z})$
terms. We plan to analyze this issue in a forthcoming publication.
Related investigations with only defect perturbations can be found
in \cite{KRW}. The bulk and defect perturbed theory is integrable
and can be solved by exploiting how the defect acts on integrable
boundaries, see \cite{BS} for the details. To summarize, in the calculation
the bootstrap relation 

\begin{equation}
T_{-}(\theta+\frac{i\pi}{3})T_{-}(\theta-\frac{i\pi}{3})=T_{-}(\theta)\label{eq:LYTboot}\end{equation}
was used together with defect unitarity and defect crossing symmetry
(\ref{Defprop}) to fix the transmission factor as \begin{equation}
T_{-}(\theta)=[b+1][b-1]\quad;\qquad[x]=i\frac{\sinh(\frac{\theta}{2}+i\frac{\pi x}{12})}{\sinh(\frac{\theta}{2}+i\frac{\pi x}{12}-i\frac{\pi}{2})}\label{eq:YLdefsol}\end{equation}
(Actually the inverse of the solution is also a solution but the two
are related by the $b\to6+b$ transformation).

We also note that the defect with parameter $b=3$ behaves as a standing
particle both from the energy and from the scattering point of view.

\subsection{Defect form factors}

In this subsection we apply the general method developed in Section
3 to determine the form factors of the defect Lee-Yang model. The
form factor can be written as \begin{equation}
F_{n}(\theta_{1},\dots,\theta_{n})=H_{n}\prod_{i}d(\theta_{i})\prod_{i<j}\frac{f_{min}(\theta_{i}-\theta_{j})}{u_{i}+u_{j}}Q_{n}(u_{1},\dots,u_{n})\label{eq:YLansatz}\end{equation}
The minimal solution of the two particle form factor equation is well-known
and reads as \cite{LY2pt}: \[
f_{min}(\theta)=\frac{u+u^{-1}-2}{u+u^{-1}+1}\, v(i\pi-\theta)\, v(-i\pi+\theta)\]
where \[
v(\theta)=\exp\left\{ 2\int_{0}^{\infty}\frac{dx}{x}e^{\frac{i\theta x}{\pi}}\frac{\sinh\frac{x}{2}\sinh\frac{x}{3}\sinh\frac{x}{6}}{\sinh^{2}x}\right\} \]
We also included the pole corresponding to the dynamical singularity
equation by the denominator. We choose the normalization of the form
factors as in the bulk \[
H_{n}=-\frac{\pi m^{2}}{4\sqrt{3}}\left(\frac{3^{\frac{1}{4}}}{2^{\frac{1}{2}}v(0)}\right)^{n}\]
$Q_{n}(u_{1},\dots,u_{n})$ is expected to be a symmetric polynomial
in $u_{i}$ and $\bar{u}_{i}$. 

Let us turn to the determination of $d(\theta)$. Due to the defect
dynamical singularity for $F_{n}(\theta_{1},\dots,\theta_{n})$ the
defect dependent term $d(\theta)$ must have a pole whenever $T_{-}(\theta)$
has a pole. Similar equation is valid for $F_{0,n}(;\theta_{1},\dots,\theta_{n})=\prod T_{-}^{-1}(\theta_{i})F_{n}(\theta_{1},\dots,\theta_{n})$
at the defect bound-states poles of $T_{+}(\theta)$ . We will take
into account that the transformation $b\leftrightarrow6-b$ exchanges
$T_{-}(\theta)$ with $T_{+}(\theta)$ and we expect that it acts
in a similar way on the form factors. (For parity invariant operators).
The minimal solution with these requirements turns out to be: \begin{equation}
d(\theta)=\frac{1}{4\sinh(\frac{\theta}{2}+\frac{i\pi}{12}(b-5))\sinh(\frac{\theta}{2}+\frac{i\pi}{12}(b-7))}=\frac{1}{\sqrt{3}+2\cos(\frac{b\pi}{6}-i\theta)}=\frac{1}{\sqrt{3}+u\nu+u^{-1}\bar{\nu}}\label{eq:YLd}\end{equation}
where we introduced $\nu=e^{i\frac{\pi b}{6}}$ and $\bar{\nu}=\nu^{-1}$.
This function satisfies two relevant relations: \[
d(\theta+i\pi)d(\theta)=\frac{1}{1-2\cos(\frac{b\pi}{3}-2i\theta)}=\frac{1}{1-u^{2}\nu^{2}-u^{-2}\bar{\nu}^{2}}\]
and \[
d(\theta+\frac{i\pi}{3})d(\theta-\frac{i\pi}{3})=\frac{1}{2\cos(\frac{b\pi}{6}-i\theta)}d(\theta)=\frac{1}{u\nu+u^{-1}\bar{\nu}}d(\theta)\]

Singularity axioms generate recursive relations between the polynomials.
The kinemetical recursion relation is given by: \[
Q_{n+2}(-u,u,u_{1},...,u_{n})=D_{n}(u,u_{1},...,u_{n})Q_{n}(u_{1},...,u_{n})\]
with\begin{eqnarray*}
D_{n}(u,u_{1},...,u_{n}) & = & (-1)^{n+1}(u^{2}\nu^{2}-1+u^{-2}\nu^{-2})\\
 &  & \frac{u}{2(\omega-\bar{\omega})}\left(\prod_{i=1}^{n}(u\omega+u_{i}\bar{\omega})(u\bar{\omega}-u_{i}\omega)-\prod_{i=1}^{n}(u\omega-u_{i}\bar{\omega})(u\bar{\omega}+u_{i}\omega)\right)\end{eqnarray*}
 where we introduced $\omega=e^{\frac{i\pi}{3}}$, $\bar{\omega}=\omega^{-1}$,
while the bound state recursion relation is :\begin{eqnarray*}
Q_{n+1}(u\omega,u\bar{\omega},u_{1},...,u_{n-1}) & = & (\nu u+\nu^{-1}u^{-1})u\prod_{i=1}^{n-1}(u+u_{i})Q_{n}(u,u_{1},...,u_{n-1})\end{eqnarray*}
Now let us try to solve these recursions.

\subsubsection{Solutions}

Since $Q_{n}(u_{1},...,u_{n})$ is supposed to be a symmetric polynomial,
it is useful to introduce the elementary symmetric polynomials $\sigma_{k}^{(n)}(u_{1},...,u_{n})$
which are defined through the generating function:\[
\prod_{i=1}^{n}(u+u_{i})=\sum_{k=0}^{n}u^{n-k}\sigma_{k}^{(n)}(u_{1},...,u_{n})\]
By means of these functions the kinemetical recursive relation for
$Q_{n}$ reads as: \[
(-1)^{n+1}Q_{n+2}(-u,u,u_{1},...,u_{n})=(u^{2}\nu^{2}-1+u^{-2}\nu^{-2})u^{2}\tilde{D}_{n}(u,u_{1},...u_{n})Q_{n}(u_{1},...,u_{n})\]
with \[
\tilde{D}_{n}(u,u_{1},...u_{n})=\sum_{k=1}^{n}\sum_{m=1,odd}^{k}\frac{\sin(\frac{2\pi}{3}m)}{\sin(\frac{2\pi}{3})}u^{2(n-k)+m}\sigma_{k}^{(n)}\sigma_{k-m}^{(n)}(-1)^{k+1}\]
We are going to find the form factors of the operators $\Phi_{\pm}(z,\bar{z})$,
$\varphi(z)$, $\bar{\varphi}(\bar{z})$ and their descendants. We
can choose $\Phi_{\pm}$ as the defect limits of the right/left bulk
fields, thus we know already all of their form factors. Taking into
account the explicit form of $d(\theta)$ together with $T_{-}(\theta)$
we find \[
Q_{1}^{\Phi_{-}}=\nu\sigma_{1}+\bar{\nu}\bar{\sigma}_{1}+\sqrt{3}\quad;\qquad Q_{1}^{\Phi_{+}}=\nu\sigma_{1}+\bar{\nu}\bar{\sigma}_{1}-\sqrt{3}\]
For the two particle form factors we get \[
Q_{2}^{\Phi_{-}}=\sigma_{1}(v^{2}\sigma_{2}+\sqrt{3}v\sigma_{1}+\sigma_{1}\bar{\sigma}_{1}+1+\sqrt{3}\bar{\nu}\bar{\sigma}_{1}+\bar{\nu}^{2}\bar{\sigma}_{2})\]
 and \[
Q_{2}^{\Phi_{+}}=\sigma_{1}(v^{2}\sigma_{2}-\sqrt{3}v\sigma_{1}+\sigma_{1}\bar{\sigma}_{1}+1-\sqrt{3}\bar{\nu}\bar{\sigma}_{1}+\bar{\nu}^{2}\bar{\sigma}_{2})\]
where we used the solution of the bulk form-factor equation $Q_{2}^{\Phi}=\sigma_{1}$.
They both satisfy the dynamical recursion relations. The asymptotics
of the solutions for $x\to\pm\infty$ reflect the dimensions of the
fields $(-\frac{1}{5},-\frac{1}{5})$.

In order to calculate the general form factor $Q_{n}$ of the fields
$\Phi_{\mp}$ we can take their bulk form factors $B_{n}^{\Phi}$
from \cite{LY2pt,DN} and rewrite it into the form (\ref{eq:YLansatz})
with the defect part given by (\ref{eq:YLd}). The resulting form
factors $Q_{n}$ are all polynomials and have the right asymptotic
properties. Similar considerations hold for the descendants of the
identity operator. However, calculating the left and right limits
they can differ. This is due to the defect perturbation and the relation
between the two must be expressed in terms of the form factors of
other fields which follows from the fact that in the UV these operators
have the same limit. The relation between the left and right limits
of the energy momentum tensor is the analogue of the defect condition
we have seen already in the case of the free boson. 

Now we would like to describe the two chiral fields $\varphi(z)$
and $\bar{\varphi}(\bar{z})$ which have dimensions $(-\frac{1}{5},0)$
and $(0,-\frac{1}{5})$. The corresponding solutions at level one
with the right asymptotics have to form\[
Q_{1}^{\varphi}=\sigma_{1}\quad;\qquad Q_{1}^{\bar{\varphi}}=\bar{\sigma}_{1}\]
They are related by the $u\leftrightarrow u^{-1}$ transformation.
Using our recursion relations we find the related solutions at level
2 \[
Q_{2}^{\varphi}=\sigma_{1}(v\sigma_{2}+\bar{\nu})\quad;\qquad Q_{2}^{\bar{\varphi}}=\bar{\sigma}_{1}(\bar{\nu}\bar{\sigma}_{2}+v)\]
Those solution are not unique as they might mix with the kernel solutions
\[
K_{1}=\sigma_{1}(\sigma_{1}\bar{\sigma}_{1}-1),\; K_{2}=\sigma_{1}(\sigma_{1}^{2}-\sigma_{2}),\, K_{3}=\sigma_{1}(\bar{\sigma}_{1}^{2}-\bar{\sigma}_{2})\]
and their descendants. Here we included the factor $\sigma_{1}$ in
all cases as $F_{2}$ must not have a dynamical pole, so the denominator
$x_{1}+x_{2}$ has to be killed. Interestingly all these kernel solutions
can be expressed in terms of $\Phi_{\mp},\,\varphi,\bar{\varphi}$
and the two descendants $\partial(\Phi_{-}-\Phi_{+}),\,\bar{\partial}(\Phi_{-}-\Phi_{+})$.
This is a nontrivial statement and shows that the form factor solutions
at level 2 having degree less then four are in one-to-one correspondence
with the operator content of the theory. 

\begin{table}
\begin{centering}
\begin{tabular}{|c|c|c|}
\hline 
Operator & $Q_{1}$ & $Q_{2}$\tabularnewline
\hline
\hline 
$\Phi_{-}$ & \emph{$\nu\sigma_{1}+\bar{\nu}\bar{\sigma}_{1}+\sqrt{3}$} & $\sigma_{1}(v^{2}\sigma_{2}+\sqrt{3}v\sigma_{1}+\sigma_{1}\bar{\sigma}_{1}+1+\sqrt{3}\bar{\nu}\bar{\sigma}_{1}+\bar{\nu}^{2}\bar{\sigma}_{2})$\tabularnewline
\hline 
$\Phi_{+}$ & \emph{$\nu\sigma_{1}+\bar{\nu}\bar{\sigma}_{1}-\sqrt{3}$} & $\sigma_{1}(v^{2}\sigma_{2}-\sqrt{3}v\sigma_{1}+\sigma_{1}\bar{\sigma}_{1}+1-\sqrt{3}\bar{\nu}\bar{\sigma}_{1}+\bar{\nu}^{2}\bar{\sigma}_{2})$\tabularnewline
\hline 
$\varphi$ & $\sigma_{1}$ & $\sigma_{1}(v\sigma_{2}+\bar{\nu})$\tabularnewline
\hline 
$\bar{\varphi}$ & $\bar{\sigma}_{1}$ & $\sigma_{1}(\bar{\nu}\bar{\sigma}_{2}+v)$\tabularnewline
\hline
\end{tabular}
\par\end{centering}

\caption{The form factor solutions of the primary fields up to level 2}

\end{table}

\subsubsection{Parity Symmetry}

In this part we analyze how the parity transformation acts on the
form factor solutions. The action of the parity operator $P$ on the
operators can be written as \[
P\mathcal{O}P^{-1}=\mathcal{O}^{P}\]
The corresponding action on the form factors is \[
P\langle0|\mathcal{O}(0)|\theta;\ \rangle=\langle0|\mathcal{O}^{P}|\ ;-\theta\rangle\]
By explicit calculations we checked that$\Phi_{-}$and $\Phi_{+}$
are parity even with \[
P\langle0|\Phi_{-}(0)|\theta;\ \rangle=\langle0|\Phi_{+}(0)|\ ;-\theta\rangle\]
 while on the contrary, $\varphi$ and $\bar{\varphi}$ are parity
odd with

\[
P\langle0|\varphi(0)|\theta;\ \rangle=-\langle0|\bar{\varphi}(0)|\ ;-\theta\rangle\]
These relations have to be confirmed in the Lagrangian framework.

\section{Boundary form factors via defects }

In this section we intend to illustrate how defects can be used to
generate new boundary form factor solutions from old ones. The underlying
fusing idea for the reflection matrices can be explained as follows:
Suppose we place an integrable defect with transmission factor $T_{-}(\theta)$
in front of an integrable boundary with reflection factor $R(\theta)$,
which satisfies unitarity and boundary crossing unitarity: \[
R(-\theta)=R^{-1}(\theta)\quad;\qquad R(\frac{i\pi}{2}-\theta)=S(2\theta)R(\frac{i\pi}{2}+\theta)\]
If we fuse the defect to the boundary the composite boundary system
will be integrable and will have reflection factor \[
\bar{R}(\theta)=T_{+}(\theta)R(\theta)T_{-}(\theta)\]
which, due to the defect unitarity and crossing equations, will satisfy
boundary unitarity and crossing unitarity. This idea has been used
to calculate the transmission factors from the already determined
reflection factors $R,\bar{R}$ in the sinh-Gordon and Lee-Yang models
in \cite{BS}. In contrast, here we would like to use the fusion idea
to generate new form factor solutions from old ones. For this purpose
we suppose that we determined already the boundary form factors $F_{n}^{\mathcal{O}}(\theta_{1},\dots,\theta_{n})$
of a boundary operator $\mathcal{O}$. It satisfies, besides the singularity
\cite{BFF} axioms, the following requirements: 

permutation \emph{\[
F_{n}^{\mathcal{O}}(\theta_{1},\dots,\theta_{i},\theta_{i+1},\dots,\theta_{n})=S(\theta_{i}-\theta_{i+1})F_{n}^{\mathcal{O}}(\theta_{1},\dots,\theta_{i+1},\theta_{i},\dots,\theta_{n})\]
}

reflection\[
F_{n}^{\mathcal{O}}(\theta_{1},\dots,\theta_{n-1},\theta_{n})=R(\theta_{n})F_{n}^{\mathcal{O}}(\theta_{1},\dots,\theta_{n-1},-\theta_{n})\]
and crossing reflection \[
F_{n}^{\mathcal{O}}(\theta_{1},\theta_{2},\dots,\theta_{n})=R(i\pi-\theta_{1})F_{n}^{\mathcal{O}}(2i\pi-\theta_{1},\theta_{2},\dots,\theta_{n})\]
We claim that the fused form factor \begin{equation}
\bar{F}_{n}^{\mathcal{O}}(\theta_{1},\dots,\theta_{n})=\prod_{i=1}^{n}T_{-}(\theta_{i})F_{n}^{\mathcal{O}}(\theta_{1},\dots,\theta_{n})\label{eq:barF}\end{equation}
 satisfies the boundary form factor axioms of the fused boundary corresponding
to the reflection factor $\bar{R}$.

Let us analyze them one by one. Since the extra factor is symmetric
in $\theta_{i}$ the permutation axiom is trivially satisfied. To
show the reflection property we use defect unitarity \begin{equation}
\bar{R}(\theta)=T_{+}(\theta)R(\theta)T_{-}(\theta)=T_{-}(-\theta)^{-1}R(\theta)T_{-}(\theta)\label{eq:Funi}\end{equation}
while for the crossing reflection we use defect crossing symmetry:
\begin{equation}
\bar{R}(i\pi-\theta)=T_{+}(i\pi-\theta)R(i\pi-\theta)T_{-}(i\pi-\theta)=T_{-}(\theta)R(i\pi-\theta)T_{-}(2i\pi-\theta)^{-1}\label{eq:Fcuni}\end{equation}
Now multiplying both sides of the reflection and crossing reflection
equation by $\prod_{i}T_{-}(\theta_{i})$ and using (\ref{eq:Funi})
and (\ref{eq:Fcuni}) the claim follows. Similarly one can show the
satisfaction of the singularity axioms \cite{BFF}. 

By this method form factor solution of a given boundary can be used
to generate form factor solutions for the fused boundary. It is practically
useful if we can follow the indentification of the operators under
the fusion procedure. This is the case for example if the operator
in the UV limit commutes with the defect. Say for example if in the
Lee-Yang model we take the form factors of the operators of the identity
module on the trivial boundary \cite{SZT}, then by the fusion procedure
we can generate the form factors of the same module on the fused $\phi$
boundary, just by multiplying the original form factor solutions with
the product of the transmission factors.

\section{Conclusion}

In the paper we initiated the form factor program for purely transmitting
integrable defect theories. We restricted our interest for a single
particle type, but the extension of the program for diagonal bulk
scatterings and diagonal transmissions is straightforward (see \cite{Ol1}
in the boundary case). We laid down axioms for the form factors of
operators localized both in the bulk and also on the defect. We determined
the solutions of the consitency requirements for bulk operators in
terms of the bulk form factors together with the transmission matrix.
These form factors determine the correlation functions of bulk operators,
which we elaborated in details for the two point functions. In the
case of defect operators we gave the general form of the solutions
and explicitly calculated for the free boson and for some operator
in the Lee-Yang model. We also described how the fusion method can
be used to generate new form factor solutions from old ones. 

In the analysis of the Lee-Yang model we observed relations between
the defect operators which should have the origin in defect conditions.
The lack of the Lagrangian definition of the model prevented us to
analyze this question. In order to acheive this aim one has to analyze
the simultaneous integrable defect and bulk perturbations of the defect
Lee-Yang model using conformal perturbation theory and establish the
relation between the bulk and defect couplings, which maintains integrability.
This approach then can be used to derive defect conditions which will
provide relations between fields living on the defect. In pushing
forward this program one has to solve the defect Lee-Yang model first.
The expilicit knowledge of the correlation functions together with
the structure constants will make it possible to bridge the operators
appearing in the form factor program to their UV counterparts by analyzing
the short distance behaviour of the two point functions obtained from
the explicit form factor solutions.

Another interesting problem is to see that we have as many polynomial
solution of the form factor equation as many local operators existing
in the theory. We have seen this coincidence in the case of the free
boson. In the case of the Lee-Yang model the nontrivial mixing between
the left and right degrees of freedom and the various cancellation
between the leading order scaling terms prevented us to perform this
analysis. Possibly a more careful analysis along the line of \cite{SZT}
would clear up this point as well. 

We have analyzed the free boson and the Lee-Yang model sofar. The
method, however, has a straightforward application for the sinh-Gordon
model adopting ideas from the boundary form factor solutions \cite{Ol2,Gabor1}. 

The defect form factors in the Lee-Yang model can be tested by extending
them for finite volume and comparing to direct TCSA data. They also
can be used to build up finite temperature defect correlation functions.
These are direct generalizatons of the related boundary analysises
developed in \cite{Gabor2,Gabor3}.

\subsection*{Acknowledgments}

We thank Laszl\'o Palla, G\'abor Tak\'acs for the useful discussions.
ZB was supported by a Bolyai Scholarship, and by OTKA K60040.

\end{document}